\documentclass[preprint,12pt,square]{elsarticle}
\usepackage{amssymb}
\usepackage[colorlinks=true]{hyperref} 
\usepackage{color}
\usepackage{multirow}
\usepackage{lineno,hyperref}
\modulolinenumbers[5]
\journal{International Journal of Refrigeration}

\begin{document}

\begin{frontmatter}

\title{Thermal Management Design and Key Technology Validation for PandaX Underground Experiment}

\author[a,b,c,d]{Tao Zhang}       
\author[a,b,c,d]{Jianglai Liu}
\author[b]{Yang Liu}
\author[b,d]{Weihao Wu}
\author[a]{Binbin Yan\corref{mycorrespondingauthorA}}
\cortext[mycorrespondingauthorA]{Corresponding author:
yanbinbin@sjtu.edu.cn}
\author[a,b,c,d]{Zhou Wang\corref{mycorrespondingauthorB}}
\cortext[mycorrespondingauthorB]{Corresponding author:
wangzhou0303@sjtu.edu.cn}

\author[b]{Zhizhen Zhou}

 \affiliation[a]{organization={New Cornerstone Science Laboratory, Tsung-Dao Lee Institute, Shanghai Jiao Tong University},
            city={Shanghai},
            postcode={201210},
             country={China}}
             
\affiliation[b]{organization={School of Physics and Astronomy, Shanghai Jiao Tong University, Key Laboratory for Particle Astrophysics and
Cosmology (MoE), Shanghai Key Laboratory for Particle Physics and Cosmology},
            city={Shanghai},
            postcode={200240},
            country={China}}

\affiliation[c]{organization={Shanghai Jiao Tong University Sichuan Research Institute},
            city={Chengdu},
            postcode={610000},
            state={Sichuan},
            country={China}}
\affiliation[d]{organization={Jinping Deep Underground Frontier Science and Dark Matter Key Laboratory of Sichuan Province},
           }

\begin{abstract}
The scale of liquid xenon experiments aimed at searching for rare events is expanding, with plans to reach up to fifty tons. These detectors and the accompanying distillation towers require a reliable cooling source that can deliver large cooling power in the liquid xenon temperature range. Previously utilized pulse tube refrigerators and GM refrigerators have the disadvantages including small cooling power, large space occupation, and non-standby mutuality, which become significant challenges to scaling up experiments. In this study, an auto-cascade refrigerator was developed using ethanol as a coolant, with improved heat transfer effect achieved by adopting a concentric shaft heat exchanger and an after-pumping heat transfer scheme. This system provides a stable cooling power of 2.5 kW at 155 K. Further, the feasibility and key technology of the centralized cooling system of 5 kW at 160 K is discussed. This study promises to simplify liquid xenon experimental auxiliary devices, thereby benefiting the PandaX-xT experiment scheme.
\end{abstract}

\begin{keyword}
\texttt Underground Experiment, Dark Matter, Rare Event,  Thermal Management, Auto-Cascade Refrigerator, Centralized Cooling System
\end{keyword}
\end{frontmatter}


\section{Introduction}
\label{Introduction}
The direct detection of dark matter is a leading particle physics experiment, with liquid xenon detectors merging as the most promising technology. The scale of the detectors has developed from the initial 10 kg to the current 10-ton level gradually, such as PandaX\cite{pandax4t}, XENON\cite{xenon10}\cite{xenon1t}, LZ(LUX)\cite{lux}. The long-term plan aims to increase this scale to fifty tons \cite{thermalmanagement},\cite{PandaX-xT},\cite{DARWIN}. Furthermore, the detection target has extended from dark matter to neutrino-less double beta decay events \cite{PandaX-xT}\cite{Xe136}. To reduce the background from cosmic rays, these experiments must operate in underground laboratories with a long construction period, high cost, and relatively limited space \cite{CJPL}. Consequently, the dimensions of the equipment underground, especially around the detector, must be carefully considered in the system design of PandaX-xT.

To improve the sensitivity of the detector, an online distillation tower for removing Rn and Kr has been developed, and a plan to integrate a cold electronic system within the detector has been made, which significantly increases the cooling power requirements. The operational temperature of the liquid xenon detector is approximately 178 K.  Liquid nitrogen is used as the powerful auxiliary cooling source for xenon injection and recovery in emergency conditions. However, pulse tube refrigerators (PTR) and GM refrigerators are commonly designed for deep temperature refrigeration below 77 K. These system is expensive with relatively small cooling power (usually below 1000 W), which makes them inadequate for the temperature requirements of the liquid xenon detector. During the design phase, it is essential to cooling power allowance, as the cooling output needs to be adjusted by heating or matching with the thermal load to achieve precise temperature control during operation \cite{cryogenics}. Consequently, different subsystems, such as the detector and the distillation tower, could not share single PTR or GM refrigerator. This leads to excessive cooling power not supporting each other mutually and the cooling power being wasted. Furthermore, liquid nitrogen is not as readily accessible at the China Jinping Underground Laboratory (CJPL), making it impractical to rely on liquid nitrogen for substantial cooling power. In conclusion, using a liquid nitrogen-assisted refrigeration scheme using PTR and GM refrigerators for small-scale liquid xenon experiments is feasible.  However, inherent limitations in cooling power and efficiency are insufficient to meet the demands of scaling up the experiments. 
 Therefore, it is essential to develop the thermal management technology for the entire liquid xenon experimental setup by adopting a centralized cooling scheme (CCS) and high cooling power refrigeration based on electric energy. This approach aims to improve the cold energy utilization efficiency, reduce the space and labor occupation of refrigeration equipment, improve the automation and reliability of the experimental system, and focus on the key task of particle detection.

Section \ref{Cooling and Heating source requirements of PandaX experiment} outlines the cooling source requirements for the PandaX experiment. Section \ref{Auto-Cascade Refrigeration (ACR)} presents the design, manufacturing, and performance testing results of the auto-cascade refrigerator (ACR). Section \ref{Design of the cryogenic centralized cooling system} discusses the design and feasibility analysis of the CCS. Section \ref{Cold Head precision temperature-controlled based on ACR (CHACR)} discusses the precision temperature control cold head based on ACR (CHACR). The discussion is concluded in section \ref{Conclusion}.
 
\section{Cooling Source Requirements of PandaX Experiment}
\label{Cooling and Heating source requirements of PandaX experiment}

In the PandaX experiments, the cooling requirements for a steady-state include the normal operation of the detector and the distillation tower, necessitating a temperature stability of 178±0.1 K\cite{cryogenics},\cite{distillation}; For transient cooling requirements, pre-cooling of xenon gas and detectors is essential, though temperature stability is not a requirement in this case \cite{thermalmanagement},\cite{liquifyandrecuperation}. During the xenon gas recovery stage before the detector ends operation requires room temperature ethanol as the heat source for liquid xenon vaporization \cite{recuperation}. Additionally, the xenon storage system requires both a cooling source and heat source \cite{thermalmanagement},\cite{liquifyandrecuperation},\cite{First-X}.
Table \ref{tab:pars} outlines the key technical parameters of the refrigerators selected for the PandaX-4T experiment\cite{cryogenics} and the PandaX-xT scheme \cite{PandaX-xT}, which need 1 kW and 3.4 kW cooling power, respectively. 

\begin{table}[b]
\renewcommand{\arraystretch}{1.5}
\centering
\caption{Cooling power requirements of the PandaX experiments}
\label{tab:pars}
\resizebox{1.0\columnwidth}{!}{

\begin{tabular}{l|l|l|l|l|l|l} 
    \hline
    \hline
    \multirow{2}{*}{Experimental device} & Cooling power & \multirow{2}{*}{Refrigerator type and model} & Lowest & \multirow{2}{*}{Output power} & Cooling Power & Electricity \\
    & requirement & & temperature &  &  at 178K & consumption   \\
    \hline
    \multirow{3}{*}{PandaX-4T Cooling Bus}  & \multirow{3}{*}{580 W} &  GM refrigerator RDK-500B (Sumitomo, Japan) & $\textless 14$ K & 80 W @ 30 K & 240 W & 6.6-6.9 kW \\
    & & Pulse tube refrigerator PT-90 (Cryomech, USA)& 32 K & 90 W  @ 80 K & 140 W & 4.3 kW \\
    & & GM refrigerator PC-150 (JEC, Japan) & 77 K & 170 W @ 165 K &  200 W  & 5.7 kW   \\ 
    \hline
    \multirow{2}{*}{PandaX-4T Distillation Tower} & 400 W & GM refrigerator AL-300 (Cryomech, USA) & 25 K & 320 W @80 K & 500 W & 7 kW \\
    & 40 W & Pulse tube refrigerator PT-60 (Cryomech, USA)& $\textless 30$ K & 60 W @ 80 K & 120 W & 2.9 kW \\
                                 
    \hline
    PandaX-xT Cooling Bus & 1300 W & GM refrigerator AL-600 (Cryomech, USA) & 25 K & 600 W @ 80 K & 950 W & 11.5 kW            \\
    \hline    
    \multirow{2}{*}{PandaX-xT Kr Distillation Tower} & 700 W & GM refrigerator KDE-300SA (Pride,China) & $\textless 30$ K & 250 W @ 77 K & 400 W & 6.6-6.9 kW \\
    & 115 W & Pulse tube refrigerator PT-60 (Cryomech, USA) & $\textless 30$ K & 60 W @ 80 K & 120 W & 2.9 kW               \\
    \hline
    PandaX-xT Rn Distillation Tower & 1300 W & GM refrigerator AL-600 (Cryomech, USA)& 25 K & 600 W @80 K & 950 W & 11.5 kW               \\
    
    \hline
    \hline
\end{tabular}}

\end{table}

\section{Auto-Cascade Refrigeration (ACR)}
\label{Auto-Cascade Refrigeration (ACR)}
The CCS requires a centralized high-power refrigeration system or cooling source. One proposed approach is to use liquid nitrogen as the primary cooling source and then control the temperature \cite{LN2}. This is feasible in ground experiments where liquid nitrogen is readily available. However, sourcing large quantities of liquid nitrogen at the CJPL poses a logistical challenge, and there is now no suitable equipment for liquid nitrogen production on site. When considering refrigeration efficiency, using liquid nitrogen to liquefy xenon gas is highly inefficient due to the significant temperature difference involved. Given that electricity is the most accessible and reliable energy at the CJPL, an electricity-based refrigerator is a more suitable option for PandaX experiments. The ACR could achieve multistage overlapping by utilizing a mixed refrigerant and a single compressor. This enables it to reach a temperature below 160 K with an ethanol coolant\cite{patent-Ref}. This facilitates effective distribution and management of cooling capacity. The 18 K temperature difference accounts for both ACR refrigeration efficiency and the feasibility of CHACR, aligning with the PandaX operating temperature of 178 K.

\subsection{ACR Prinple and Design}
\label{Refrigeration technical requirement and scheme design}
Based on the above discussion, using two ACR units in operation with one as a backup can balance cost-effectiveness and reliability. Consequently, the cooling power of a single ACR design should exceed 2 kW at 160 K.

\begin{figure}[htbp]
    \centering\includegraphics[width=13cm, angle=0]{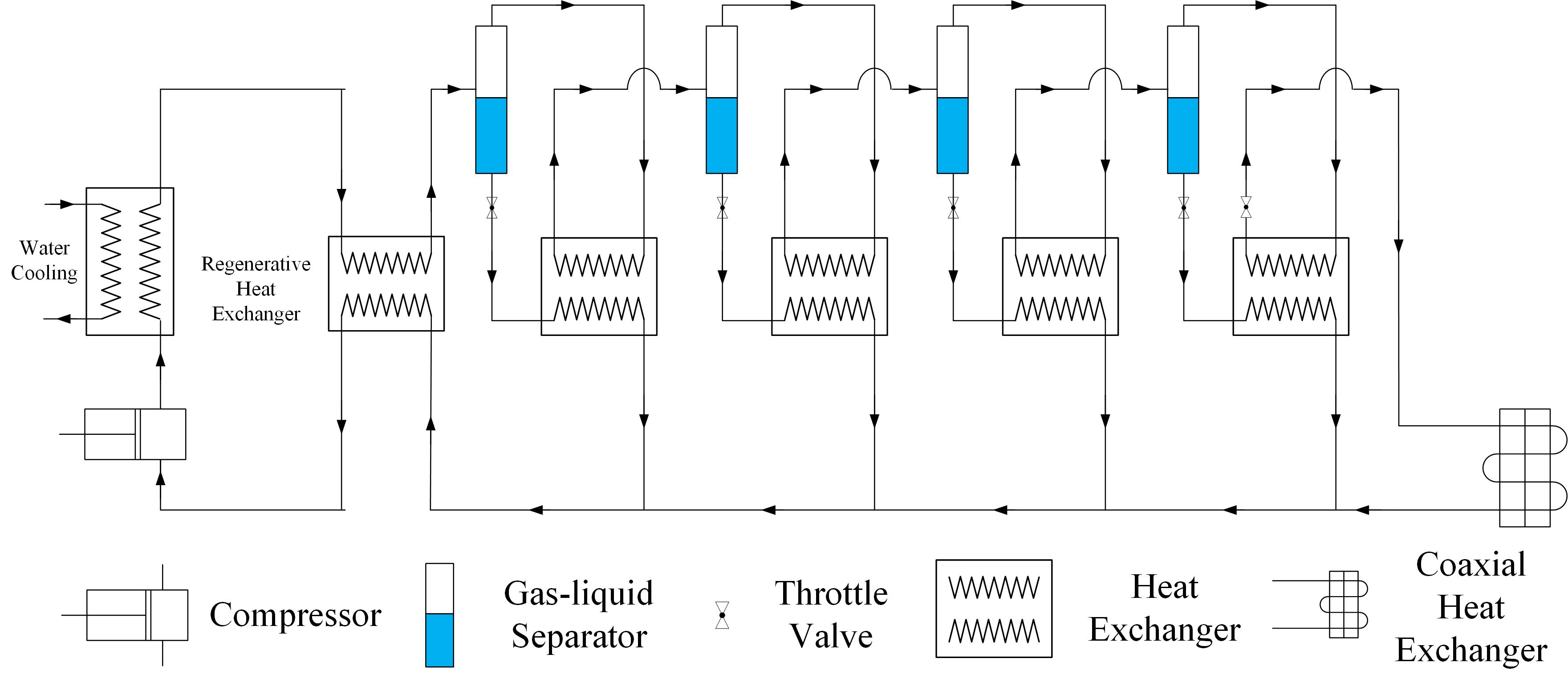}
    \caption{Working principle of ACR}
    \label{fig:Principle}
    \end{figure}

The ACR unit employs a mixed refrigerant and a single compressor to facilitate a multi-stage cascade, achieving an evaporation temperature of approximately 150 K. The operation principle is illustrated in figure \ref{fig:Principle}. Initially, the mixed refrigerant is compressed into a high-temperature and high-pressure state by the compressor. It then undergoes cooling through a water-cooled condenser and a  regenerative heat exchanger, resulting in the liquefaction of some high boiling point components. In contrast, the low boiling point components remain gaseous. Then the mixed refrigerant of the gas-liquid phase enters the first stage gas-liquid separator, and the liquid part enters the first stage intermediate heat exchanger after being depressurized and cooled via a throttle valve. This process further cools the low boiling point gas phase components from the gas-liquid separator. The two-phase mixed refrigerant then flows into the subsequent gas-liquid separator. This sequence is repeated until the component with the lowest boiling point is completely liquefied in the fourth stage intermediate heat exchanger. Subsequently, it was throttled and evaporated, to achieve a temperature of 150 K. Afterwards, it entered the final stage evaporator to cool the flowing coolant\cite{patent-Ref} . The gaseous refrigerant exiting the intermediate heat exchanger at each is compressed and pressurized by the compressor after flowing through the heat exchanger, thereby completing the refrigeration cycle. The ACR could achieve effective refrigeration output below 160 K, which greatly enhances distribution management and simplifies the cooling system. Compared with the detector operating temperature of 178 K, the 18 K temperature difference margin balances the refrigeration efficiency and realizes the precision temperature control modules based on heat conduction. 

\begin{figure}[htbp]
    \centering\includegraphics[width=13cm, angle=0]{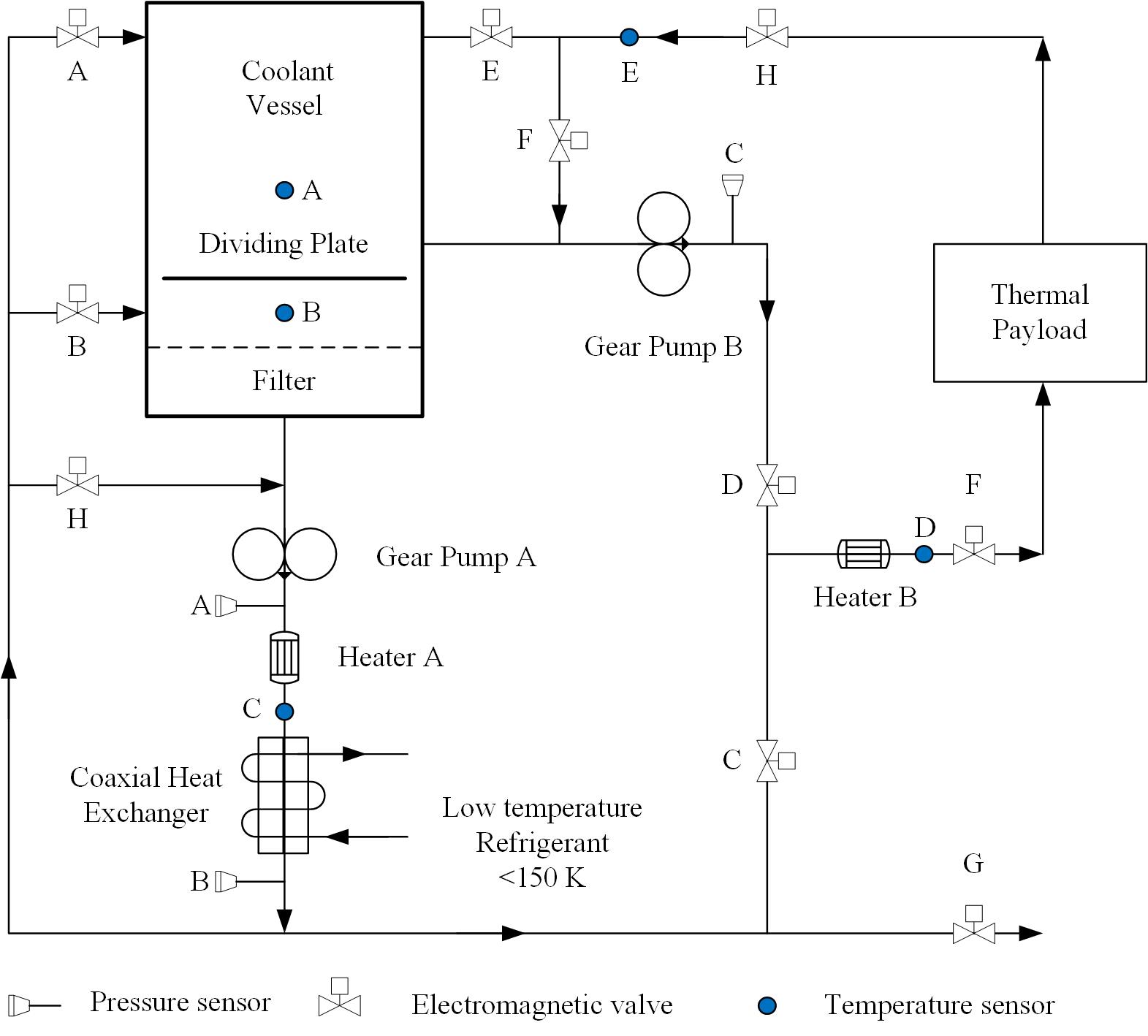}
    \caption{Coolant circulation diagram of ACR}
    \label{fig:design}
    \end{figure}

The coolant circulation principle of the ACR is illustrated in figure~\ref{fig:design}. The ethanol coolant is driven by gear pump A (Model NP1700) from the coolant vessel to the coaxial heat exchanger, which function as the refrigerator's evaporator for cooling. After cooling process, the coolant either returns to the vessel or is directly transferred to the external thermal load. The total volume of the coolant vessel is about 50 L, which is divided by a plate into two parts of 40 L and 10 L with temperature sensors A and B, respectively. There is a gap of about 1 cm between the dividing plate and the inner wall for coolant flow. A filter of 400 meshes is installed at the bottom of the vessel to protect gear pump A. Gear pump A is driven by a servo motor, facilitating continuous adjustment of the rotary speed, boasting a flow rate of 30 L/min at 2000 r/min. An electric heater A is positioned between gear pump A and the coaxial heat exchanger to aid in ice melting. The upper vessel is connected to gear pump B, enabling the coolant to be pumped out to the external thermal load. Furthermore, an electric heater B is installed at the output pipeline, utilized during heating mode. The ACR working principle is described in detail, combined with various operational modes as follows.

\subsubsection{Pre-cooling Mode}
\label{Pre-cooling Mode}
Cooling the coolant from room temperature to 160 K in the coolant vessel takes approximately 2 h. To expedite this process, the ACR can start pre-cooling in this mode. The ACR is rated for a temperature of 160 K, resulting in a significant compressor load when the coolant flowing through the evaporator is at a much higher temperature. The deviation from the rated condition shortens the lifespan of ACR. To address this issue, the design incorporates a two-step cooling process: first, the coolant in the lower vessel is cooled, followed by a gradual cooling of the coolant in the upper vessel.

For instance, the ACR begins at room temperature with the default operation parameters. During the pre-cooling mode, gear pump A delays its activation for a few seconds, and the coolant temperature in the lower vessel determines the flow rate. The flow rates are 5 L/min, 10 L/min, 20 L/min and 30 L/min corresponding to the temperature higher than 273 K, from 223 K to 273 K, from 173 K to 223 K, and lower than 173 K, respectively. If the coolant temperature in the lower vessel is above 173 K, the coolant cooled by the evaporator enters the lower vessel via valve B to be further cooled to below 173 K soon because of the smaller volume with the smaller heat capacity of the lower vessel. Subsequently, valve A opens, allowing the cryogenic coolant to flow into the upper vessel, gradually lowering the temperature of its internal coolant. Simultaneously, the coolant with a higher temperature in the upper vessel flows into the lower vessel via the gap around the dividing plate and mixes with the low-temperature coolant, causing the temperature to increase. Once the temperature exceeds 173 K, valve A closes while valve B opens to cool down the lower vessel's coolant preferentially. During the pre-cooling mode, the status of valves A and B switches several times, and the temperature of the upper vessel drops to 173 K gradually. Afterward, the stable cooling process starts with valve A opening and valve B closing until the coolant reaches the desired setting temperature.

\subsubsection{Indirect output Mode}
\label{Indirect output Mode}
This operation mode delivers coolant to effectively cool down the thermal load, functioning as an upgraded pre-cooling mode. Once the indirect output mode starts, the gear pumps A and B are switched on simultaneously, and the cooling process is consistent with the pre-cooling mode. Valves D and F open while pump B starts, and the thermal load temperature is measured by temperature sensor E after 1 min. Suppose the temperature difference between sensors A and E exceeds 30 K, valve E opens intermittently to reduce the thermal stress, and the temperature difference determines the opening duty rate. The cold coolant in the upper vessel is then mixed with the reflux coolant and delivered to the thermal load by gear pump B for cooling purposes. When the temperature difference between sensors A and E decreases to the predetermined value, such as 10 K, valve E opens while valve F closes. This adjustment enhances heat exchange between the cold coolant in the upper vessel and the thermal load, resulting in a stable cooling output for the thermal load. Simultaneously, the refrigerator continues to cool the coolant in the vessel, and valve C opens when the temperatures of the upper vessel and the reflux flow drop to 173 K. Valves A, B, and D, and gear pump B, are closed to cool down the thermal load directly with the coldest coolant from the evaporator, improving cooling efficiency. In this mode, stability during the thermal load cooling process is maintained, preventing any adverse effects of the high temperature thermal load on the lifespan of the refrigerator.

{\subsubsection{Direct output Mode}
\label{Direct output Mode}
In this mode, gear pump A is activated, causing valves A, B, D, F, and H to close while valves C and E open. Following a 1 min delay, the compressor begins the cooling process, regardless of the current temperatures of the coolant and conditions of the thermal load. This mode allows the coldest coolant to flow directly to the thermal load facilitating faster cooling than the indirect mode. This mode is specifically designed for situations requiring rapid cooling.}

\subsubsection{Deicing Mode}
\label{Deicing Mode}
The ethanol in the evaporator has the risk of being solidified when the temperature gets close to or lower than its freezing point, for the designed operating temperature of the refrigerator is low. Once the ethanol solidifies in the evaporator, the heat transfer deteriorates, with the flow resistance increasing and the flow rate decreasing, further worsening the heat transfer. The heat exchanger enters a positive feedback state, rapidly losing its cooling capacity. Two pressure sensors are installed before and after the evaporator for monitoring, and the evaporator is considered frozen when the pressure difference exceeds 4 bar. Subsequently, the ACR enters the deicing mode automatically. When the ACR operates in the indirect output mode, the gear pump B, valves D and E open, while valves C and F close, and the thermal payload is still cooling by the coolant in the upper vessel. When the ACR operates in pre-cooling mode, the valves A and B close, and the valve H opens. After the deicing mode starts, valves A and B close while valve H opens, and the electric heater of 5 kW is switched on, then the flow rate of gear pump A reduces to 2 L/min. The temperature of the heated ethanol rises above 190 K with a volume of approximately 5 L in the evaporator. This is sufficient to melt all solid ethanol within 3 min in deicing mode, thereby restoring the heat exchange function of the evaporator. The deicing mode automatically exits after 3 min, returning the system back to normal operation.

\subsubsection{Heating Mode}
\label{Heating Mode}
The PandaX detector requires heat sources for three key processes: vacuum pumping process, xenon gas recuperation, and the Photomultiplier Tube (PMT) cryogenic test rewarming process. Once the heating mode starts, valves D and F, gear pump B, and electric heater B are triggered. Heat is transferred from the ethanol refrigerant to the load until it reaches the set temperature, and the system automatically shuts down. During the heating mode, the circulating coolant bypasses the vessel and evaporator to save energy. The maximum heating temperature is set at 333 K, considering the maximum operating temperature of the detector and the boiling point of the ethanol, corresponding to the ethanol vapor pressure of 47 kPa. Proper sealing of the closed circulation tube line is essential to maintain safety. Following the heating and vacuuming process of the detector, the cooling process is preferred to utilize the indirect output mode, given the large temperature difference between the detector and the coolant, ensuring a smooth the the cooling operation.

\subsection{Coolant}
\label{coolant}
The cooling system can be categorized into direct cooling and indirect cooling based on the positional relationship between the evaporator and the thermal load when implementing the cold source. Direct cooling involves the heat exchange between the cold refrigerant and the thermal load within the evaporator. This method offers higher heat exchange efficiency, however, it lacks flexibility. In contrast, indirect cooling uses a coolant as an intermediate medium. In this system, the evaporator first cools the coolant, which is then circulated to the thermal load for cooling. This coolant is  then referred to as the secondary refrigerant. Indirect cooling simplifies the operation, maintenance, and management of the refrigeration system, facilitating a more flexible distribution and control of the cooling power. Moreover, this system often operates with fewer larger refrigeration units, making it more efficient and reliable. Additionally, the cooling temperature is stable and reliable because the coolant is also the cold storage medium with a large thermal capacity or high thermal inertia. However, it is worth noting that indirect cooling may lead to a reduction in refrigeration efficiency compared to direct cooling, as it involves an additional stage of heat exchange.

The coolant must remain in the liquid phase during circulation between the evaporator and the thermal load. Anhydrous ethanol is used as the coolant, with melting point and boiling point of 159 K and 352 K, respectively, and the vapor pressure is 8 kPa at 298 K. The volatility of the ethanol is moderate without stains left after volatilizing hindering the pollution towards the experimental equipment. The risk is that ethanol is flammable, with a flash point of 282 K and a spontaneous combustion point of 695 K.  Therefore ensuring the sealing reliability of the ACR is vital. To maintain safety, it is necessary to enhance ventilation and install ethanol sensors to monitor potential leaks.

Generally, melting point measurements are conducted under slow heating, cooling and static state \cite{Chemistry}. The flow rate of the ethanol coolant in the ACR is substantial, occurring in a highly turbulent state with a large cooling rate and temperature gradient. Therefore, the liquid flow can be sustained even under sub-cooling conditions, potentially for two reasons: First, the solid ethanol exhibits weak strength when the temperature is slightly lower than the melting point, akin to the negative correlation of the temperature and the strength of the sea ice and metal materials\cite{seaice},\cite{melting}. Secondly, the ethanol coolant is in a fluid-ice state, a mixture of liquid ethanol and tiny solid ethanol crystals \cite{crystallization}. When the temperature is close to or slightly lower than the freezing point, the viscosity of ethanol significantly increases, increasing the flow resistance and changing the flow characteristics. In short, the cooling capacity is increased. In the static test, the ethanol at 150 K contained within stainless steel containers has a viscosity similar to honey. Its non-freezing property in the over-cooling state makes ethanol a suitable coolant, with the lowest output temperature of the ACR recorded at about 150 K during testing.

\subsection{Evaporator and Heat Exchanger}
\label{Evaporator and heatexchanger}
\begin{figure}[htbp]
    \centering\includegraphics[width=13cm, angle=0]{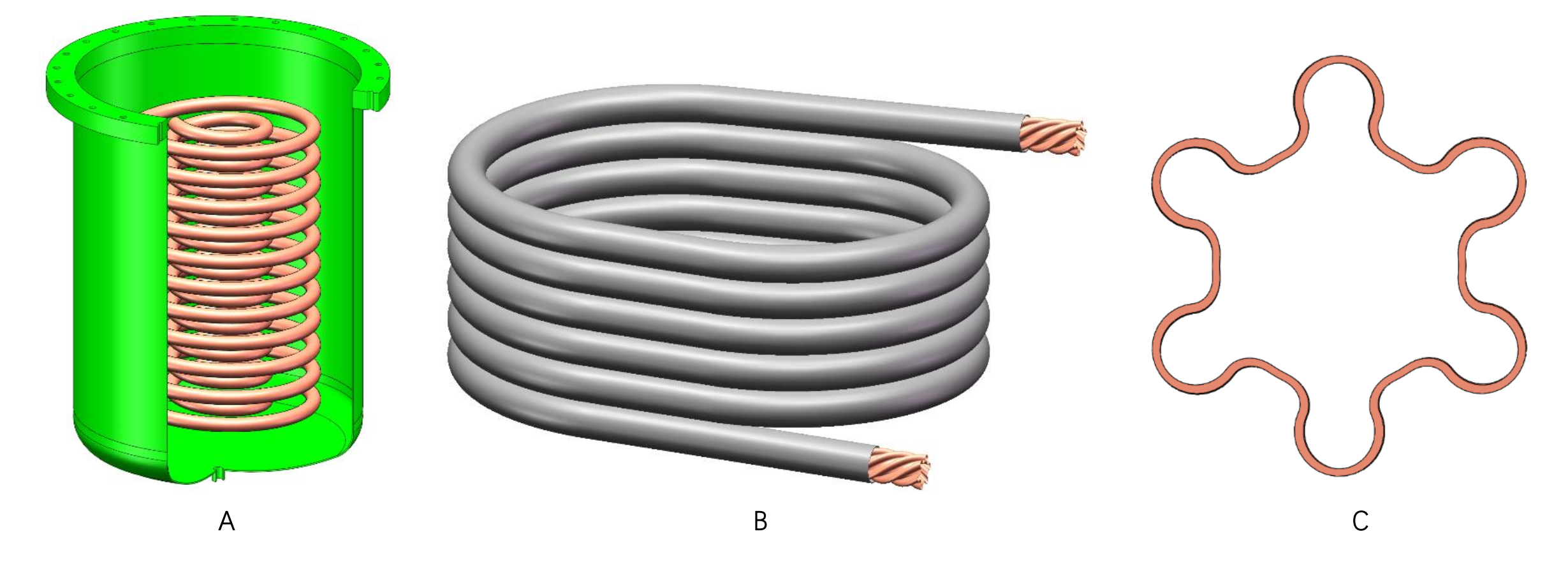}
    \caption{ The drawings of coil heat exchanger and coaxial heat exchanger with its cross-section form}
    \label{fig:cross_section}
    \end{figure}

The primary technical indicators of a cryogenic refrigerator include its output cooling power and the lowest attainable temperature, both of which are constrained by the performance of the heat exchanger. An increase in cooling power is essential to compensate for the efficiency reduction of the heat exchanger, and the acquisition cost of the cooling power gets higher as the working temperature decreases, so the requirement of the heat exchange efficiency is strict \cite{cryoheatexchanger}. In the previously utilized refrigerator, the coil heat exchanger is the evaporator within the ethanol coolant vessel, as illustrated in figure~\ref{fig:cross_section}A. The refrigerant flows inside the coil while the ethanol coolant is outside. Due to the large cross-sectional area of the vessel and the compact size of the coil, the flow rate of the coolant is relatively slow during circulation. When the working temperature is much higher than the freezing point of the ethanol, its low viscosity causes the ascendant heat exchange effect. However, as the temperature decreases, the viscosity of the ethanol increases, which in turn thickens the velocity and temperature boundary layers on the coil surface increase. This transition causes the flow state to change from turbulent to laminar, deteriorating the heat exchange effect. Thus, the lowest output temperature of the refrigerator is limited.

The plate heat exchanger is a highly efficient and compact heat exchanger comprising a series of parallel thin metal plates with corrugated surfaces. This design is considered as the heat exchanger with a collection of multiple narrow sub-flow channels in parallel with periodic section changes, facilitating high fluid turbulence and effective heat exchange  \cite{heatexchanger}. However, once the coolant begins to freeze in a sub-channel near its freezing point, the flow resistance in that sub-channel increases, leading to a decrease in flow rate. This facilitates the solidification process until the sub-channel is blocked and loses the heat exchange ability, and may leave only one effective sub-channel finally. Consequently, the effective heat exchange area is far less than the original design, making the plate heat exchanger unsuitable for such conditions.

The coaxial heat exchanger is a concentric tube comprising inner and outer tubes of different diameters coiled into round or oblong shapes, as illustrated in figure~\ref{fig:Principle}, figure~\ref{fig:design} and figure~\ref{fig:cross_section}B. Compared to the coil heat exchanger, shell and tube heat exchanger, and plate heat exchanger, the casing heat exchanger has only one section of the limited flow channel. The cross-section of the inner tube is a star-shaped coolant channel, as illustrated in figure~\ref{fig:cross_section}C. Generally, the ethanol coolant is in a highly turbulent state with high heat exchange efficiency due to the uneven and section-changing tube surface. Because the high-speed turbulence flow strongly impacts the tube wall, the solidified micro-crystals are difficult to deposit and grow. However, this design has the drawback of relatively high flow resistance, necessitating the application of a circulating pump with high-pressure output \cite{cryoheatexchanger}. The gear pump can generate an output pressure of 2 MPa and a maximum suction pressure of 0.09 MPa. The flow resistance of the evaporator increases when the temperature is close to the freezing point of the ethanol. Therefore, the gear pump needs to be installed before the evaporator, allowing the cooling effect of the heat exchanger to fully operate and mitigate the negative impact of pump-induced heating of the coolant. Therefore, the ACR could achieve effective output at lower temperatures.

\subsection{Circulation Pump}
\label{Circulation Pump}
An impeller centrifugal pump is typically employed in this kind of refrigerator due to its simple structure and high efficiency. Its operating principle revolves around transferring energy to the fluid medium via a rotating impeller. However, when the viscosity of the medium increases, the shaft power increases, and the efficiency and the flow decrease. The loss of mechanical energy converted into heat, elevating the medium temperature, and the output flow is significantly affected by the outlet pressure \cite{centrifugalpump}. Previous tests revealed that increasing the power of the centrifugal pump could not improve the output cooling power of the refrigerator because its self-heating effect offsets the cooling ability enhanced by increasing the pump power. In contrast, the gear pump is a positive displacement pump, which pressurizes the medium by interlocking the gears, making it capable of handling fluids with higher viscosity. Its output flow rate is nearly proportional to the rotary speed, almost without being influenced by the outlet pressure \cite{displacementpump}. Because the rated temperature of the refrigerator is close to the freezing point of ethanol, which has varying viscosity at room temperature, the gear pump is as the preferred choice for the circulating pump. Furthermore, a servo motor is used to adjust the speed of the pump, featuring a soft start to ensure the reliability of the refrigerator. The rotation acceleration is adjustable up to a maximum of 5 $r/s^2$.  A pressure sensor is installed at the output of the pump to maintain the pressure below 0.6 MPa. Should the pressure exceed 0.8 MPa, the pump will shut down and trigger an alarm.

\subsection{Test Results and Analysis}
\label{Test results analysis}
\begin{figure}[htbp]
    \centering\includegraphics[width=14cm, angle=0]{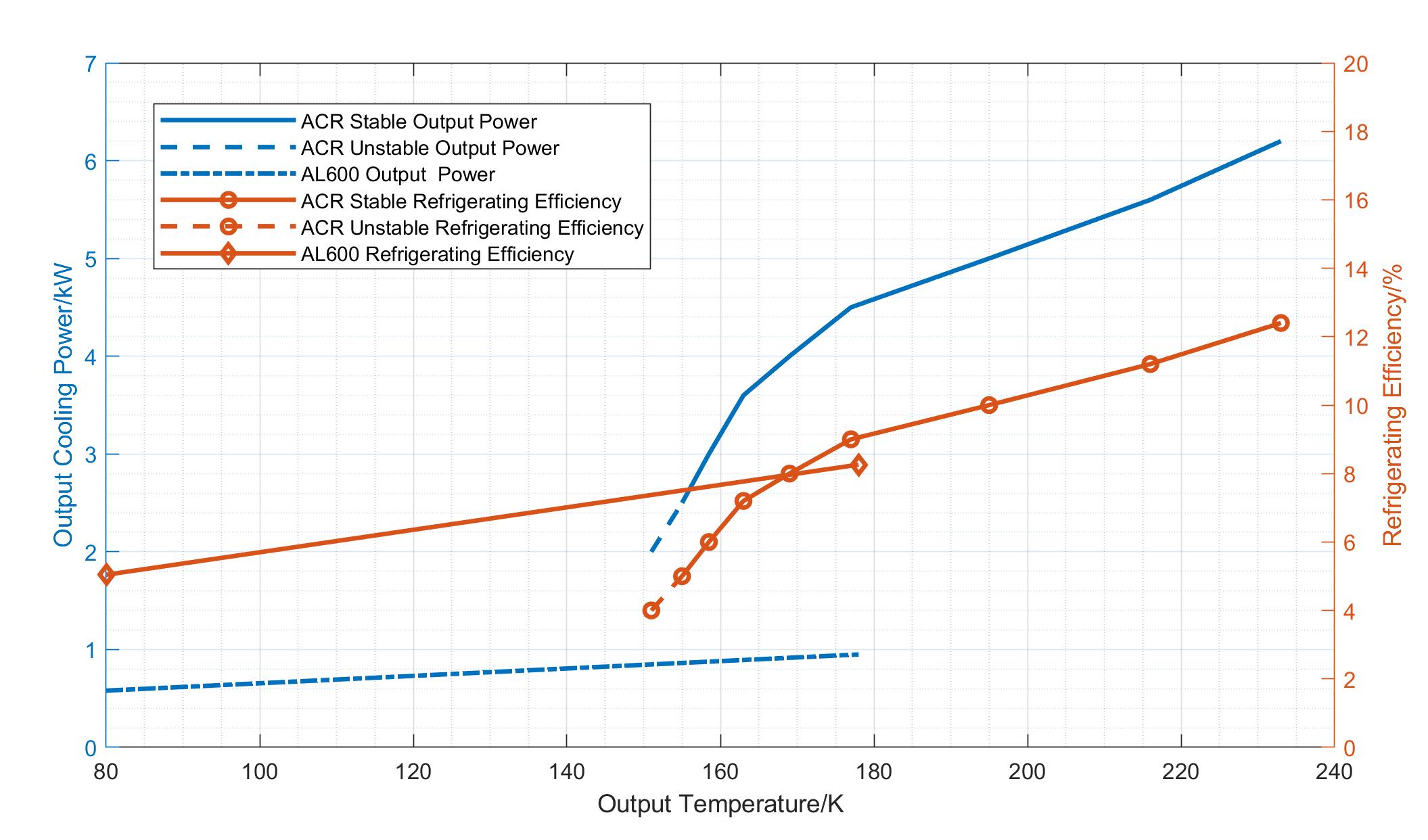}
    \caption{The output performance of ACR and AL600}
    \label{fig:Output_ACR}
    \end{figure}

The ACR dimensions are 2 m length, 1.25 m width and 1.9 m height with a compressor that has an electric power of about 45 kW. Following the development of the ACR, the electrical heater B in figure~\ref{fig:design} serves as the virtual thermal load for testing purposes. The ACR peak power consumption is about 50 kw. The lowest output temperature achieved by the ACR is 151 K, corresponding to an effective output power of 2 kw. However, this temperature is not stable enough for condensing the coolant at the heat exchanger. The ACR maintains a stable output temperature of 155 K with an output cooling power of 2.5 kW. Additionally, the cooling power is 3 kW, 3.6 kW, 4 kW, 4.5 kW, 5 kW, 5.6 kW and 6.2 kW, respectively, corresponding to temperatures of 158.5 K, 163 K, 169 K, 177 K, 195 K, 216 K and 233 K. The performance curve shown in figure~\ref{fig:Output_ACR}, indicates that the ACR cooling power exceeds that of the PTR and GM refrigerators listed in table~\ref{tab:pars}.

As illustrated in figure~\ref{fig:Output_ACR}, the refrigeration efficiency of the ACR is near the AL600 refrigerator around 170 K. However, its price per unit of cooling power is significantly lower than AL600. In the applications of PTR and GM refrigerators, the excess cooling power is compensated by electrical heating, and the effective output cooling power is below the rated value. The coolant of the ACR has a large heat capacity, and the compressor stops when the coolant temperature is below the setting temperature. The rest power of the circulation pump and the control system power is much smaller than the compressor. When accurate temperature output is required, the precision temperature control method described in section~\ref{Cold Head precision temperature-controlled based on ACR (CHACR)} would be used in the ACR to lose a part of the temperature difference, reflecting the decline of refrigeration efficiency. In a word, the refrigeration efficiency of ACR, PTR, and GM refrigerators can not be compared directly. It is more reasonable to be considered comprehensively in the specific system.

\begin{figure}[htbp]
    \centering\includegraphics[width=8cm, angle=0]{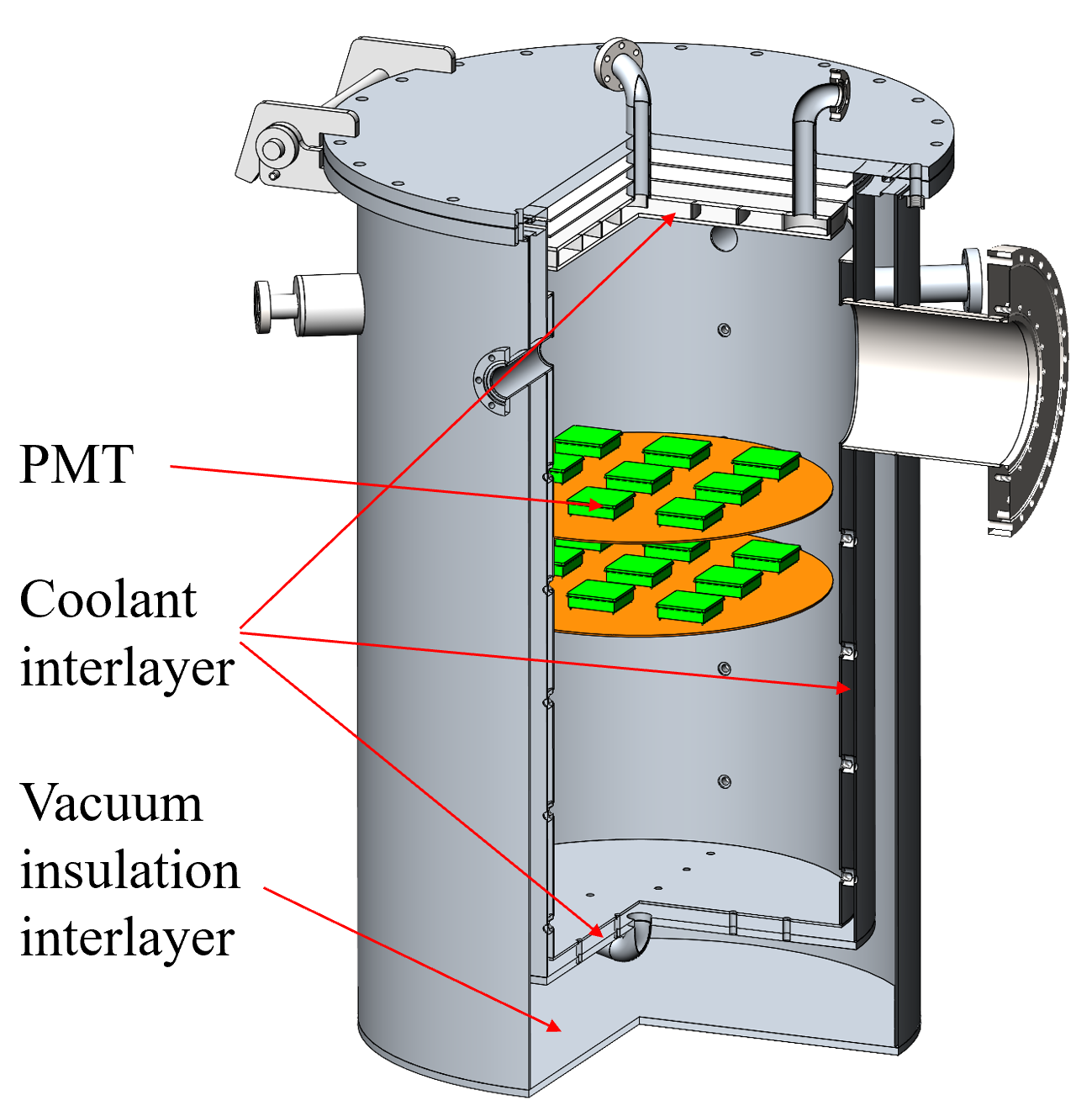}
    \caption{The structure of dewar for PMT cryogenic test}
    \label{fig:PMT_test_dewar}
    \end{figure}
  
\begin{figure}[htbp]
    \centering\includegraphics[width=15cm, angle=0]{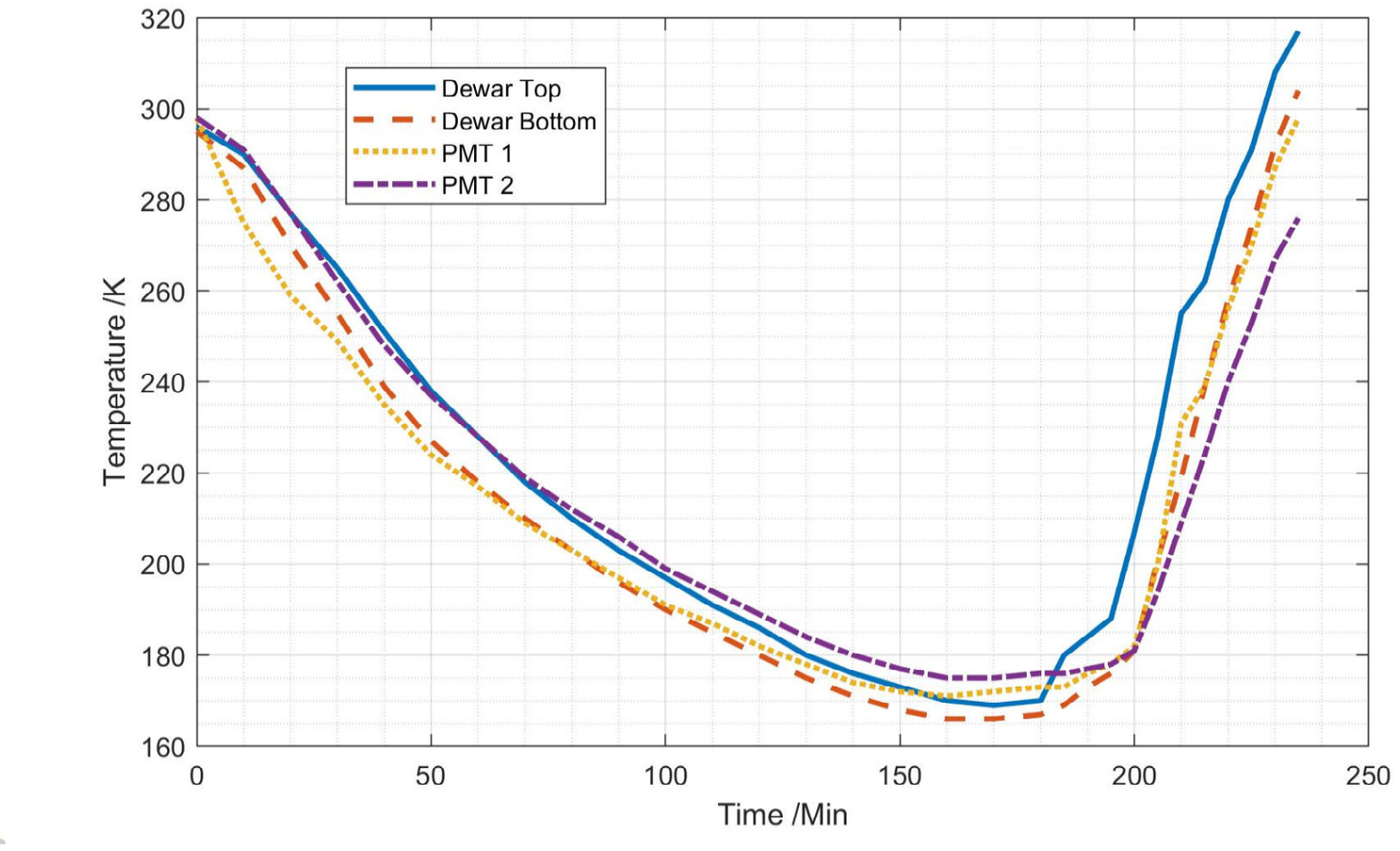}
    \caption{The temperature response of PMT cryogenic test dewar}
    \label{fig:cryogenic_test}
    \end{figure}

The dewar for the PMT cryogenic test is illustrated in figure~\ref{fig:PMT_test_dewar}, with 450 mm inner diameters and 735 mm height. The total weight of the stainless steel interlayer in contact with the coolant is about 80 kg. The ACR and dewar are connected via a vacuum-insulated hose with a maximum coolant flow rate of 20 L/min. Temperatures are measured at various locations: the gas at the top and bottom of the dewar, and the PMTs. The cooling process, utilizing direct output cooling mode, takes about 150 min to lower the temperature from 300 K to 170 K. In contrast, the return to room temperature via heating mode requires approximately 40 min, as illustrated in figure~\ref{fig:cryogenic_test}. The heat transfers from the inner wall of the dewar to the PMTs depends on air, leading to large thermal resistance that introduces a noticeable time delay in the cooling and heating process.

\section{Design of Centralized Cooling System (CCS)}
\label{Design of the cryogenic centralized cooling system}
As mentioned before, the PandaX experiment requires cooling at multiple locations, CCS is a promising solution. Cold coolant is transported to various thermal payloads via vacuum-insulated pipelines, with the ACR serving as the primary cold source and liquid nitrogen as the back cold source. Similarly, large-scale scientific facilities such as the Beijing Electron-Positron Collider, Large Hadron Collider, China Spallation Neutron Source, and Large Proton Collider have adopted or plan to adopt the CCS with liquid helium temperature zone that provides several kilowatts of cooling power over distances of several kilometers \cite{development}. It should be noted that ethanol coolant is flammable, necessitating using helium mass spectrometry for leak detection in CCS pipelines.

\subsection{Vacuum-insulated Pipeline design}
\label{Pipeline design}
The length and width of the CJPL-II experimental hall, where the PandaX experiment is located, are about 60 m and 14 m, respectively. To ensure cooling coverage distance is designed to reach 100 m, resulting in a total cooling pipeline length of about 200 m. The inner diameter of the pipeline is 50 mm with vacuum insulation and CF flange connection. The maximum flow rate of the pipeline is about 300 L/min by design. Based on the design of a round-trip pipeline temperature difference of 2 K, the maximum cooling power can reach 20 kW, meeting all the cooling requirements of the future experimental system.
 The inner pipeline and the insulation layer need to pass the helium mass spectrometry leak detection test for safety requirements.

The CCS 2000 L of ethanol coolant has a large heat capacity and cold storage ability to stabilize the temperature and enhance reliability. The heat leakage of the pipeline is about 1 W/m, and the heat leakage of each valve is about 1W. Therefore, the total heat leakage power is below 300 W, a level that is acceptable considering the CCS can handle several kilowatts.

Moreover, room-temperature ethanol is needed as a heat source for xenon recuperation \cite{recuperation} and the operation of the distillation tower. Due to the temperature difference being at 140 K, the pipe diameter of 20 mm is enough. 

\subsection{Multiple Cooling Sources Management}
\label{Refrigerator access and reserve}
\begin{figure}[htbp]
    \centering\includegraphics[width=13cm, angle=0]{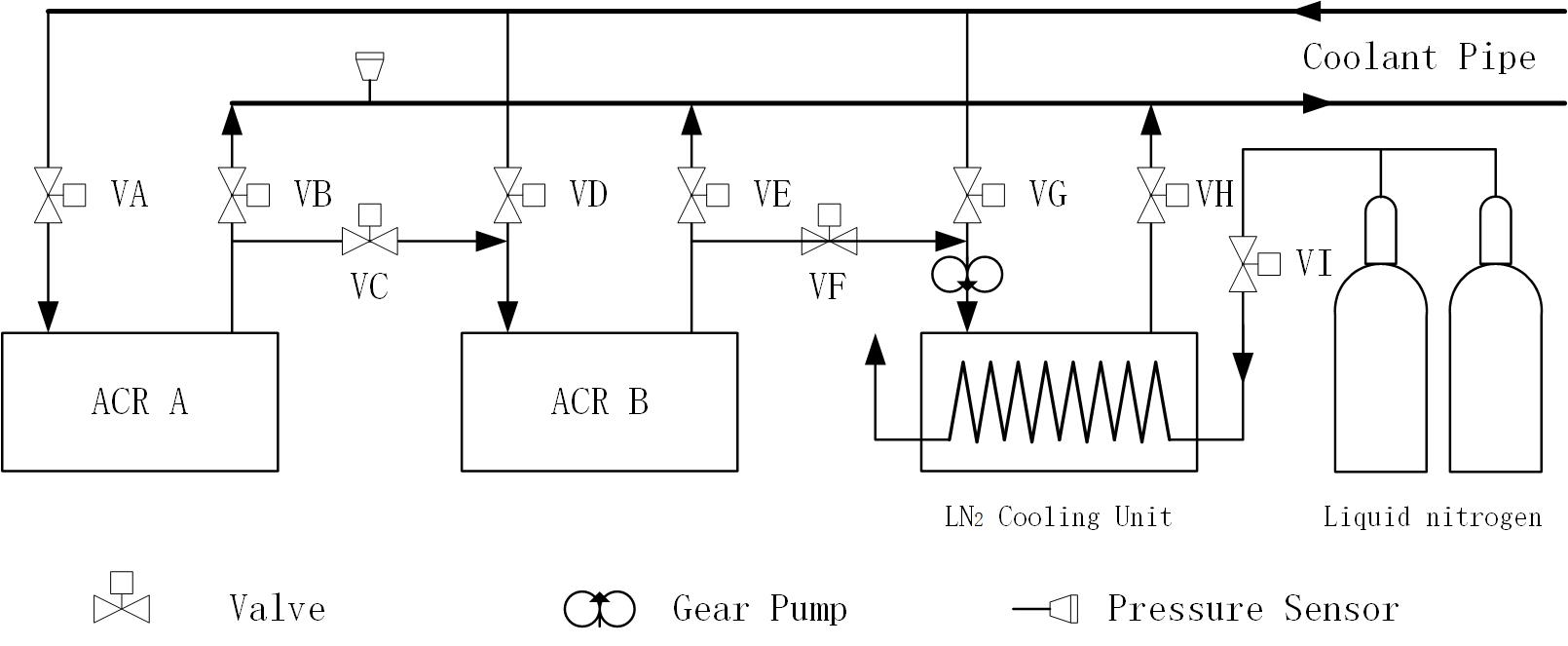}
    \caption{Multiple cooling source management}
    \label{fig:cooling_supply}
    \end{figure}

To ensure the reliability of the cooling supply and to meet the high-power cooling requirements during stages such as detector pre-cooling and liquid xenon injection, multiple ACRs need to operate simultaneously. Additionally, a backup ACR that can be activated at any time is required. Besides the backup ACR, a liquid nitrogen cooling unit is also required to improve the reliability of the system during prolonged power failures or other extreme conditions. The strategy for managing multiple cooling sources is illustrated in figure~\ref{fig:cooling_supply}.
    
Taking PandaX-4T as an example, a single ACR can meet all the cooling power requirements under normal operation. In this case, ACR A in figure~\ref{fig:cooling_supply} is running, valves VA and VB open, while valve VC closes, allowing the coolant circulate. Meanwhile, ACR B and the liquid nitrogen cooling unit are connected to the CCS as thermal loads with valves VD, VE, VG, and VH open, to enable slow circulation of the coolant, maintaining a low-temperature state and ensuring that cooling power could be delivered at any moment. However, in cases where the thermal load is relatively larger, such as PanaX-xT, a single ACR may not provide sufficient power; hence, it may be necessary to operate two or more ACRs in conjunction with the liquid nitrogen cooling unit.

The CCS contains about 2000 L coolant, with a heat capacity of about 4$\times 10^{6}$ J/K, When the ACRs are temporarily shut down, the accumulated cooling capacity can be used to maintain the normal operation of the system. During this shutdown, the coolant temperature increases by about 1 K per 800 seconds under a thermal load of 5 kW. This duration is adequate to start the diesel generator as an emergency backup power source or liquid nitrogen cooling unit. The amount of available liquid nitrogen is about 160 L per container of 200 L, which can provide about 6 kWh of cooling power with the 200 J/g vaporization heat. Unlike the PTR and GM refrigerators, a high power and large capacity UPS is not necessary for ACR; a smaller UPS is sufficient to power the controller and pump, leading to cost and space saving for the experimental system, while improving overall reliability.

\subsection{Typical Thermal Load and Access Strategy Analysis}
\label{Representative load and access analysis}
The ACR is relatively cost-effective and offer substantial cooling power, which can meet most of the cooling requirements of xenon detectors. Four typical thermal loads were analyzed, using PandaX-4T detector as case study, provide a simplified thermal management solution to illustrate design approach and provide a reference for future system designs.

\subsubsection{Thermal Load with Precision Temperature Control}
\label{Precision temperature control of thermal load}
\begin{figure}[htbp]
    \centering\includegraphics[width=13cm, angle=0]{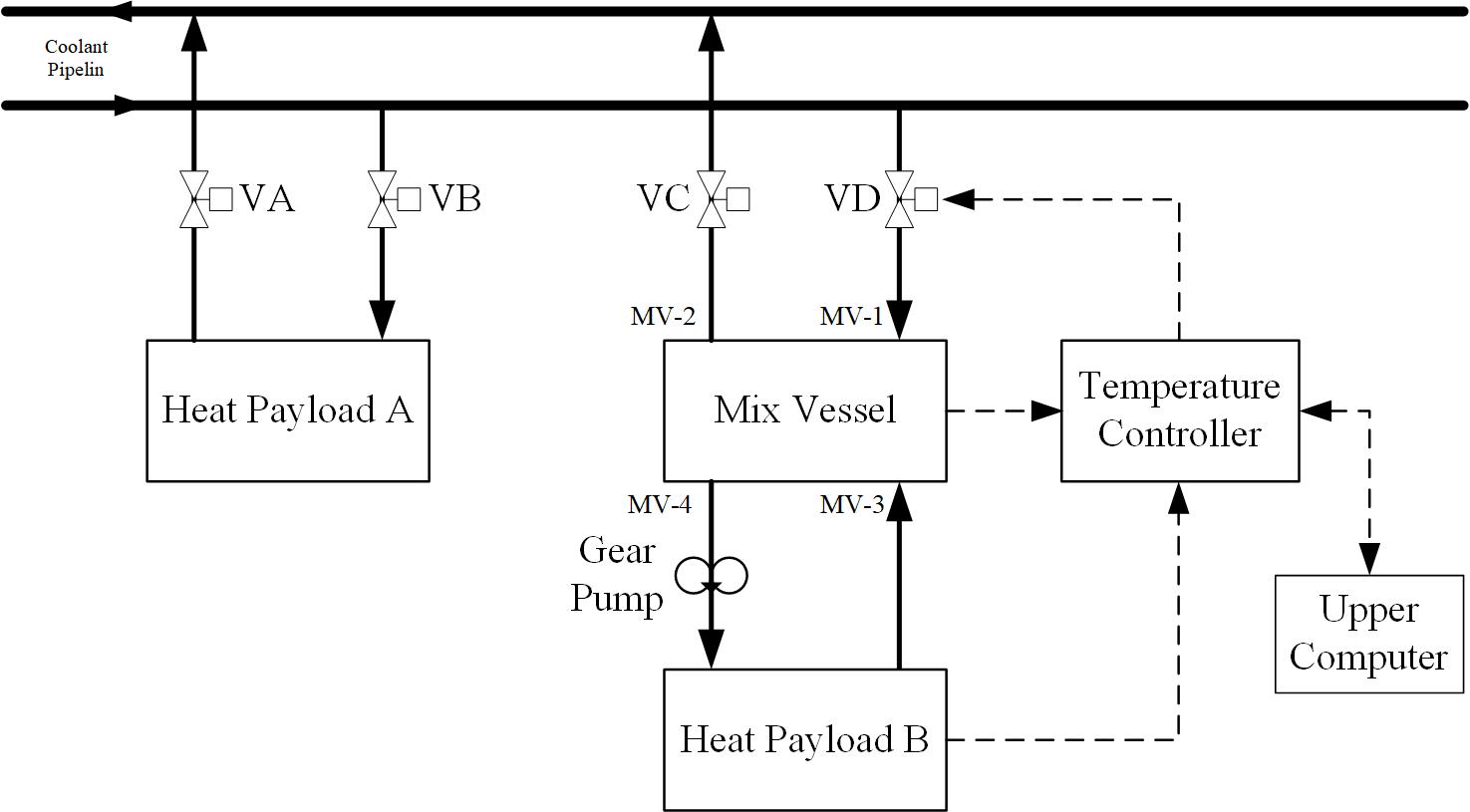}
    \caption{Thermal load access design of CCS}
    \label{fig:thermal_load}
    \end{figure}

The cooling power requirements are not large when the PandaX detector and the distillation tower are in 178 K steady-state operation. However, the temperature stability requirement is strict, with ±0.1 K or even ±0.01 K tolerance. The “thermal load A” in figure~\ref{fig:thermal_load} can be directly connected to the CCS. The design of the precision temperature control cold head based on the ACR is described in section \ref{Cold Head precision temperature-controlled based on ACR (CHACR)}, where the low-temperature coolant is the primary cooling source. 

\subsubsection{High power thermal load with high temperature}
\label{High power thermal load with high temperature}
Typical thermal loads include the initial cooling of a large-scale xenon gas storage system and the PandaX detector, which generally which typically requires several days to transition from room temperature to the target temperature. A large cooling power is required, regardless of temperature stability. All the ACRs should be in operation. Maintaining temperature stability in the coolant is critical when other thermal loads with precise temperature control requirements are active within the CCS. A mixer vessel connecting these thermal loads to the cooling pipeline is beneficial, as illustrated in figure~\ref{fig:thermal_load}. 

The mixer is a transition between thermal loads B and the CCS. Functionally, the mixer is similar to the coolant storage tank on a refrigerator. The mixer has four ports for coolant input and output. The coolant driven by a gear pump flows out from port MV-4 at a relatively high flow rate to cool the “thermal load B” and then returns to the mixer through port MV-3. The low-temperature coolant enters the mixer through valve VD and port MV-1 and then returns to the CCS main coolant pipeline through valve VC and port MV-2. Ports MV-1 and MV-3 positioned closely and aligned to achieve a better mixing effect to improve temperature uniformity. The temperature signal is input to the temperature controller, which calculates the opening ratio of the control valve VD based on the upper computer command and the established control law, ensuring that the coolant in the mixer remains within the designated temperature range. Although the temperature of the coolant returning to the pipeline through valve VC may be high, the temperature stability of the cooling supply system can still be maintained when the returning flow rate is low. For instance, the Rn adsorption unit requires a cooling source of 180$\sim$200 K \cite{LN2}, which can be integrated with the CCS through the mixer.

\subsubsection{{ Thermal Load of Xenon Purification Process}}
\label{Thermal load of xenon purification cycling}
\begin{figure}[hp]
    \centering\includegraphics[width=15cm, angle=0]{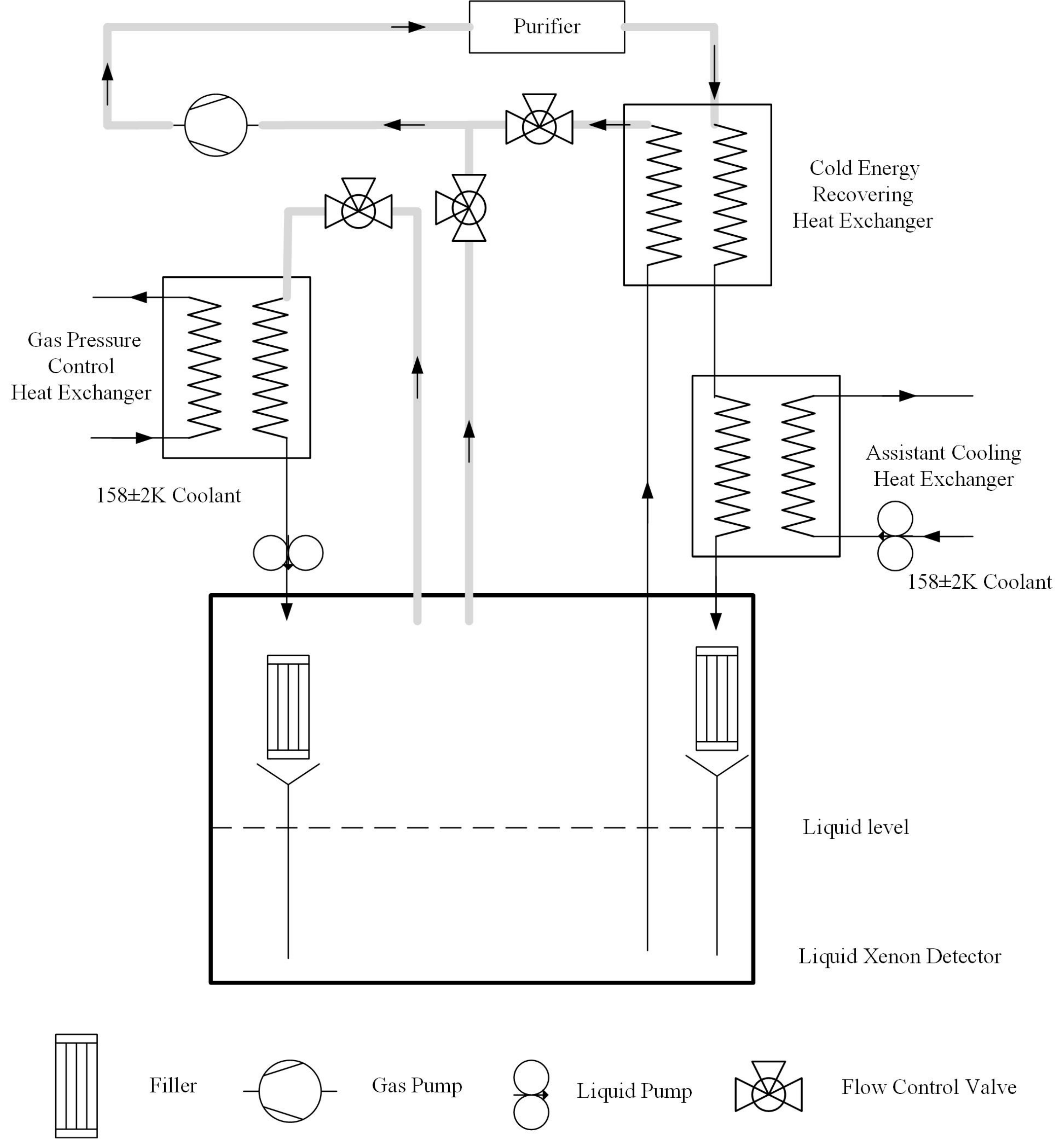}
    \caption{Design diagram of the cooling supplement}
    \label{fig:cooling_replenishment}
    \end{figure}

One of the main thermal loads during the normal operation of the detector is proportional to the circulation flow rate and the heat exchange coefficient of the circulation system. At a circulation flow rate of 155 L/min, the heat exchange efficiency for the PandaX-4T detector is approximately 97.5$\%$ \cite{cryogenics}, resulting in a thermal load of approximately 45 W. It was effectively managed by the precision temperature-controlled cold head. For future detectors, the thermal load introduced by the circulation and purification process can be calculated.  As illustrated in figure~\ref{fig:cooling_replenishment}, if assistant secondary cooling heat exchanger is added after the cold energy recovering heat exchanger with the ethanol coolant at 163±2 K on the cold side, the purified and cold-recovered gas-liquid mixture is liquefied fully and fed into the detector. This process helps mitigate the thermal load associated with xenon circulation and purification, preventing any unwanted heating of the detector. The specific heat capacity of liquid xenon at 0.2 MPa is about 0.34 kJ/(kg·K). 

Using the PandaX-4T detector as a case study for calculation, the cooling power derived from the coolant of 16 K lower than the detector reaches up to 60 W. With the maximum thermal load previously estimated at 45 W, this results in a potential reduction of 105 W in cooling power requirements for the other cooling systems associated with the detector. The temperature fluctuation of ±2 K corresponds to a cooling power fluctuation of ±8 W. In contrast, the variations in thermal load fluctuation due to heat exchange efficiency are minimal in comparison to the circulation rate. By adjusting the circulation flow rate of the coolant, the heat exchange power of the auxiliary cooling heat exchanger can be modified, necessitating adjustments in the input cooling power to the detector. The densities of liquid xenon at 178 K and 163 K are 2854 $kg/m^3$ and 2956 $kg/m^3$, respectively. The 3.6$\%$ difference in density does not affect the performance of the detector. Therefore, this circulation and purification process provides adjustable cooling power to the detector, potentially reducing the cooling power requirement of the cooling bus, to the point where it might even be eliminated.

\subsubsection{Thermal Load of Pressure Precision Control of the Detector}
\label{Pressure control of the detector}
During the operation of a dual-phase detector, the significant impact of thermal load is the pressure increase in the detector. The current cooling systems used in PandaX transfer cooling power from the cold head to the gas xenon through copper cold fingers, enabling the liquification of xenon gas to reduce the detector pressure. The temperature of the cold head is strictly controlled to ensure temperature stability throughout the process \cite{cryogenics}. Furthermore, the pressure within the detector can be adjusted by changing the gas phase volume and the total amount of xenon present \cite{patent}. To facilitate this, a gas pressure control heat exchanger can be set up based on the convenient supply of coolant, as illustrated in figure~\ref{fig:cooling_replenishment}, the coolant is at the cold side, while the hot side connects to the gas within the detector via pipeline. This setup functions similarly to a low-temperature adsorption pump functionally, and the gaseous xenon is absorbed and liquefied, then accumulated to the bottom for gravity and transported to the detector due to gravity or a gear pump. Considering the temperature gradient during heat transfer, the coolant of 163 K ensures an equilibrium pressure of 0.1 MPa on the hot side of the heat exchanger. When compared to the rated operating pressure of 0.2 MPa in the detector, the requirement of the flow resistance in the gas pipeline is not strict for the pressure difference of 0.1 MPa. The flow control valve on the gas pipeline plays a crucial role in precisely managing the gas flow from the detector to the pressure-controlled heat exchanger, ensuring stable pressure within the detector. For instance, the steady-state operation of PandaX-4T requires a cooling power of about 100 W and the xenon vaporization heat is about 92 J/g. The pressure in the detector can be balanced by liquefying 1 g/s xenon gas in the heat exchanger. This method is the direct measurement and control of the detector pressure \cite{patent}, which can control the detector pressure more quickly and accurately and reduce the technical requirements of the temperature control compared to the precise temperature control method using the cold head \cite{cryogenics}. The gas phase of the detector can also be connected to the inlet of the gas circulation pump via the gas pipeline and flow control valve, thereby enabling precise pressure control function of the detector. To maintain the liquid xenon temperature entering the detector constant, a packing structure similar to a distillation column is added to the return liquid xenon pipeline. The packing is located in the gas phase of the detector, and the liquid xenon with lower temperature exchanges heat and mass with the gas xenon on the surface of the packing; some gas xenon is liquefied and quickly balanced to the equilibrium temperature, then injected into the specified position of the detector by gravity, which can keep the stability of the detector operation without the requirement of the complex temperature control system. The packing acts as an adsorption pump for the gaseous xenon.

\section{Design of Cold Head based on ACR (CHACR)}
\label{Cold Head precision temperature-controlled based on ACR (CHACR)}
\begin{figure}[h]
    \centering\includegraphics[width=10cm, angle=0]{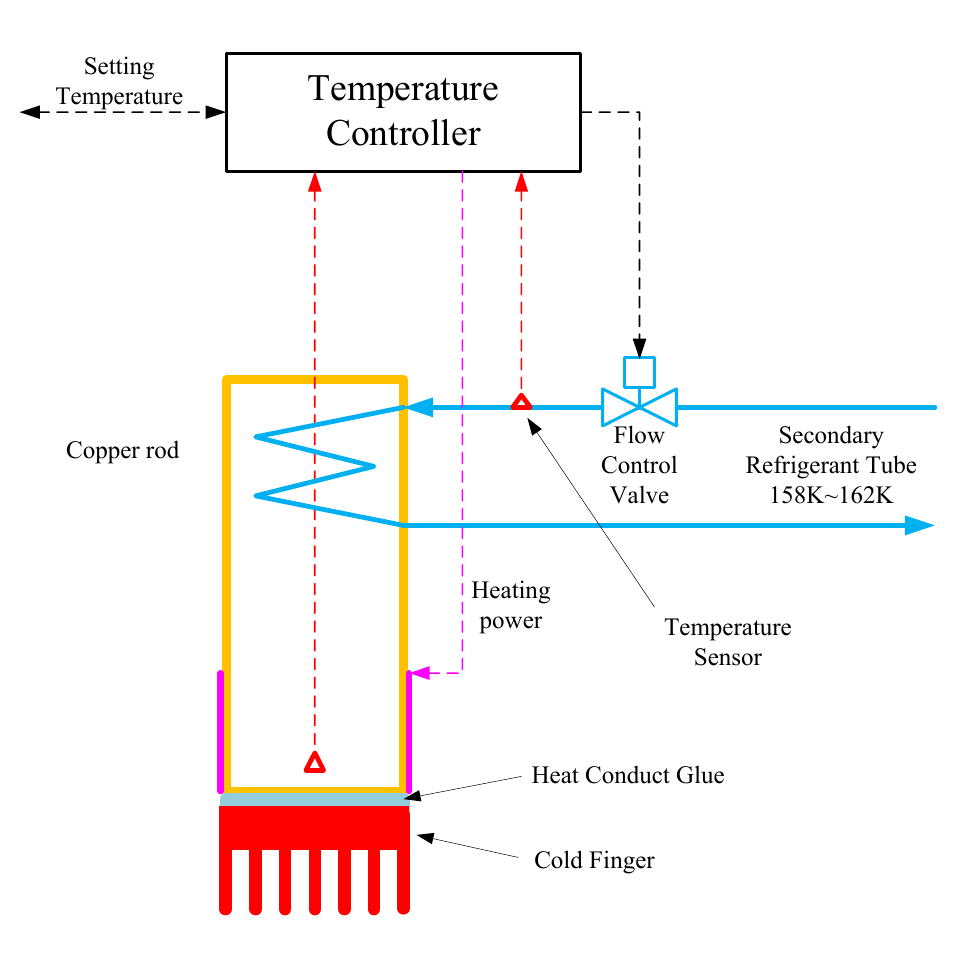}
    \caption{The design principle of the CHACR}
    \label{fig:temperature_control}
    \end{figure}

The ACR can achieve a temperature below 160 K. However, its operating principles determine the temperature fluctuation of the output is relatively large at the level of ±2 K, which cannot meet the temperature stability requirement of 178±0.1 K for the PandaX experiment system. It is feasible to design and manufacture CHACR concerning PTR and GM cold heads. The coolant flow control based on a flow regulating valve is used as the coarse adjustment, precise adjustment is carried out by electric heating, and the control signal is all from the temperature controller. Figure~\ref{fig:temperature_control} shows the design principle of CHACR.

The coolant channels are designed at the top of the copper rod with a diameter of 150 mm, a precision temperature sensor is at the bottom of the copper rod for temperature control, and the temperature is defined as the output temperature of CHACR. The bottom of the copper rod is thermally connected to the cold finger via the thermal adhesive, and the cold finger can cool down the xenon gas in the PandaX detector or the distillation tower directly. For stable thermal load, increasing the cold coolant flow rate can decrease the output temperature of the CHACR and vice versa. As a thermally conductive structure, the copper rod has a low-pass filtering effect on temperature fluctuations at the input end. Therefore a 10 Hz sampling frequency is sufficient for the CHACR temperature controller. The coolant flow rate should be operated within 10\% to 30\% of the full precision electric heating power of the temperature controller, which can balance system efficiency and fast response to transient thermal loads. The cooling power of the AL600 cold head with a 127 mm diameter is about 950 W at 178 K. The CHACR can obtain the same cooling power or more.

\section{Conclusion}
\label{Conclusion}
The developed ACR uses the gear pump to drive ethanol coolant and utilizes the coaxial heat exchanger after the pump to achieve the stable output of 2.5 kW at 155 K, which can meet the temperature and power requirements of the liquid xenon experiments. Further, a centralized cooling system with a precision temperature control cold head based on ACR is designed, which is convenient for distributing cooling capacity to improve the utilization efficiency and reduce the complexity of the auxiliary devices of the liquid xenon experimental system. The technology can be used in the PandaX-xT and provide references for the design and construction of underground laboratory infrastructures.

\section*{Acknowledgments}

This work was supported by grants from the Ministry of Science and Technology of China (No. 2023YFA1606203, No. 2023YFA1606200), Shanghai Pilot Program for Basic Research — Shanghai Jiao Tong University (No. 21TQ1400218), the Science and Technology Commission of Shanghai Municipality (Grant No.22JC1410100), National Science Foundation of China (No. 12090060, U23B2070).We thank for the support by the Fundamental Research Funds for the Central Universities. We also thank the sponsorship from the Chinese Academy of Sciences Center for Excellence in Particle Physics (CCEPP), Hongwen Foundation in Hong Kong, New Cornerstone Science Foundation, Tencent Foundation in China, and Yangyang Development Fund, and the Yalong River Hydropower Development Company Ltd. for indispensable logistical support and other help. Finally, the authors thank the technical support from Qiangxing LIN of Shanghai SUOFU Co., Ltd. and Yonghuang DENG of Shenzhen Ruixue Refrigeration Equipment Co., Ltd, we have united and cooperated, making a small push for the development of productive force.

\bibliography{main.bbl}

\begin{thebibliography}{10}
\expandafter\ifx\csname url\endcsname\relax
  \def\url#1{\texttt{#1}}\fi
\expandafter\ifx\csname urlprefix\endcsname\relax\def\urlprefix{URL }\fi
\expandafter\ifx\csname href\endcsname\relax
  \def\href#1#2{#2} \def\path#1{#1}\fi

\bibitem{pandax4t}
H.~Zhang, A.~{Abdukerim}, W.~Chen, et. {al}, Dark matter direct search
  sensitivity of the pandax-4t experiment, Science China Physics, Mechanics \&
  Astronomy 62~(3) (2019) 031011.
\newblock \href {http://dx.doi.org/10.1007/s11433-018-9259-0}
  {\path{doi:10.1007/s11433-018-9259-0}}.

\bibitem{xenon10}
J.~Angle, E.~{Aprile}, F.~{Arneodo}, et. {al}, First results from the xenon10
  dark matter experiment at the gran sasso national laboratory, Physical Review
  Letters 100 (2008) 021303.
\newblock \href {http://dx.doi.org/10.1103/PhysRevLett.100.021303}
  {\path{doi:10.1103/PhysRevLett.100.021303}}.

\bibitem{xenon1t}
E.~Aprile, J.~{Aalbers}, F.~Agostini, et. {al}, The xenon 1t dark matter
  experiment, The European Physical Journal C 77~(12) (2017) 881.
\newblock \href {http://dx.doi.org/10.1140/epjc/s10052-017-5326-3}
  {\path{doi:10.1140/epjc/s10052-017-5326-3}}.

\bibitem{lux}
D.~Akerib, X.~{Bai}, S.~{Bedikian}, et. {al}, The large underground xenon (lux)
  experiment, Nuclear Instruments and Methods in Physics Research A 704 (2013)
  111--126.
\newblock \href {http://dx.doi.org/10.1016/j.nima.2012.11.135}
  {\path{doi:10.1016/j.nima.2012.11.135}}.

\bibitem{thermalmanagement}
T.~Zhang, J.~{Liu}, High purity xenon management and heat management design for
  hundred tons scale liquid xenon experiment, Vacuum and Cryogenics 29~(2)
  (2023) 200--208.
\newblock \href {http://dx.doi.org/10.3969/j.issn.1006-7086.2023.02.014}
  {\path{doi:10.3969/j.issn.1006-7086.2023.02.014}}.

\bibitem{PandaX-xT}
A.~Abdukerim, Z.~Bo, W.~Chen, et. {al}, Pandax-xt – a deep underground
  multi-ten-tonne liquid xenon observatory, Sci. China-Phys. Mech.Astron.
  68~(2) (2025) 221011.
\newblock \href {http://arxiv.org/abs/2402.03596} {\path{arXiv:2402.03596}},
  \href {http://dx.doi.org/https://doi.org/10.1007/s11433-024-2539-y}
  {\path{doi:https://doi.org/10.1007/s11433-024-2539-y}}.

\bibitem{DARWIN}
L.~Baudis, Darwin/xlzd: A future xenon observatory for dark matter and other
  rare interactions, Nuclear Physics B 1003 (2024) 116473.
\newblock \href
  {http://dx.doi.org/https://doi.org/10.1016/j.nuclphysb.2024.116473}
  {\path{doi:https://doi.org/10.1016/j.nuclphysb.2024.116473}}.

\bibitem{Xe136}
K.~Ni, Y.~{Lai}, A.~{Abdusalam}, et. {al}, Searching for neutrino-less double
  beta decay of 136xe with pandax-ii liquid xenon detector, Chinese Physics C
  43~(11) (2009) 113001.
\newblock \href {http://dx.doi.org/10.1088/1674-1137/43/11/113001}
  {\path{doi:10.1088/1674-1137/43/11/113001}}.

\bibitem{CJPL}
J.~Li, X.~{Ji}, W.~{Haxton}, et. {al}, The second-phase development of the
  china jinping underground laboratory, Physics Procedia 61 (2015) 576--585.
\newblock \href {http://dx.doi.org/10.1016/j.phpro.2014.12.055}
  {\path{doi:10.1016/j.phpro.2014.12.055}}.

\bibitem{cryogenics}
L.~Zhao, X.~{Cui}, W.~{Ma}, et. {al}, The cryogenics and xenon handling system
  for the pandax-4t experiment, Journal of Instrumentation 16 (2021) T06007.
\newblock \href {http://dx.doi.org/10.1088/1748-0221/16/06/T06007}
  {\path{doi:10.1088/1748-0221/16/06/T06007}}.

\bibitem{distillation}
R.~Yan, Z.~{Wang}, X.~{Cui}, et. {al}, Pandax-4t cryogenic distillation system
  for removing krypton from xenon, Review of Scientific Instruments 92 (2021)
  123303.
\newblock \href {http://dx.doi.org/10.1063/5.0065154}
  {\path{doi:10.1063/5.0065154}}.

\bibitem{liquifyandrecuperation}
T.~Zhang, J.~{Liu}, Large scale storage of high purity gas with rapid
  liquefaction and recovery device, Chinese Patent: CN115307050A.

\bibitem{recuperation}
Z.~Wang, W.~{Ma}, T.~{Zhang}, et. {al}, Design and operation of the pandax-4t
  high speed ultra-high purity xenon recuperation system, Journal of
  Instrumentation 17 (2022) T10008.
\newblock \href {http://dx.doi.org/10.1088/1748-0221/17/10/T10008}
  {\path{doi:10.1088/1748-0221/17/10/T10008}}.

\bibitem{First-X}
X.~Wang, Z.~Lei, Y.~Ju, et. {al}, Design, construction and commissioning of the
  pandax-30t liquid xenon management system, Journal of Instrumentation 18~(05)
  (2023) P05028.
\newblock \href {http://dx.doi.org/10.1088/1748-0221/18/05/P05028}
  {\path{doi:10.1088/1748-0221/18/05/P05028}}.

\bibitem{LN2}
X.~Jiang, L.~{Zhao}, S.~{Wang}, et. {al}, A $ln_2$-based cryogenic system
  prototype for future pandax experiment, Journal of Instrumentation 17 (2022)
  T07005.
\newblock \href {http://dx.doi.org/10.1088/1748-0221/17/07/T07005}
  {\path{doi:10.1088/1748-0221/17/07/T07005}}.

\bibitem{patent-Ref}
T.~Zhang, J.~{Liu}, A wide temperature range low-temperature refrigerator,
  Chinese Patent 202320362287.7.

\bibitem{Chemistry}
P.~Bruice, Essential organic chemistry 3rd, New York: Pearson.

\bibitem{seaice}
D.~Meng, X.~{Chen}, S.~{Ji}, Prediction and analysis of flexural strength of
  sea ice based on recurrent neural network, Mechanics in Engineering 44~(3)
  (2022) 580--589.
\newblock \href {http://dx.doi.org/10.6052/1000-0879-21-434}
  {\path{doi:10.6052/1000-0879-21-434}}.

\bibitem{melting}
T.~Li, M.~{Hassan}, K.~{Kuwana}, Performance of secondary aluminum melting:
  Thermodynamic analysis and plant-site experiments, Energy 31 (2006)
  1769--1779.
\newblock \href {http://dx.doi.org/10.1016/j.energy.2005.08.005}
  {\path{doi:10.1016/j.energy.2005.08.005}}.

\bibitem{crystallization}
X.~Shao, C.~{Wang}, Y.~{Yang}, et. {al}, Screening of sugar alcohols and their
  binary eutectic mixtures as phase change materials for low-to-medium
  temperature latent heat storage: Non- isothermal melting and crystallization
  behaviors, Energy 160 (2018) 1078--1090.
\newblock \href {http://dx.doi.org/10.1016/j.energy.2018.07.081}
  {\path{doi:10.1016/j.energy.2018.07.081}}.

\bibitem{cryoheatexchanger}
D.~Popov, K.~{Fikiin}, B.~{Stankov}, et. {al}, Cryogenic heat exchangers for
  process cooling and renewable energy storage: A review, Applied Thermal
  Engineering 153 (2019) 275--290.
\newblock \href {http://dx.doi.org/10.1016/j.applthermaleng.2019.02.106}
  {\path{doi:10.1016/j.applthermaleng.2019.02.106}}.

\bibitem{heatexchanger}
O.~Arsenyeva, J.~{Kleme$\check{s}$}, E.~{Klochock}, et. {al}, The effect of
  plate size and corrugation pattern on plate heat exchanger performance in
  specific conditions of steam-air mixture condensation, Energy 263 (2023)
  125958.
\newblock \href {http://dx.doi.org/10.1016/j.energy.2022.125958}
  {\path{doi:10.1016/j.energy.2022.125958}}.

\bibitem{centrifugalpump}
L.~Zhang, X.~{Wang}, P.~{Wang}, et. {al}, Optimization of a centrifugal pump to
  improve hydraulic efficiency and reduce hydro-induced vibration, Energy 268
  (2023) 126677.
\newblock \href {http://dx.doi.org/10.1016/j.energy.2023.126677}
  {\path{doi:10.1016/j.energy.2023.126677}}.

\bibitem{displacementpump}
A.~Josifovic, J.~{Roberts}, J.~{Corney}, et. {al}, Reducing the environmental
  impact of hydraulic fracturing through design optimisation of positive
  displacement pumps, Energy 115 (2016) 1216--1233.
\newblock \href {http://dx.doi.org/10.1016/j.energy.2016.09.016}
  {\path{doi:10.1016/j.energy.2016.09.016}}.

\bibitem{development}
Y.~Zhang, G.~{Wang}, Z.~{Hu}, et. {al}, Review of cryogenic development and
  advance research in large scientific projects, Cryogenics 3 (2016) 17--22.
\newblock \href {http://dx.doi.org/1000-6516(2016)03-0017-06}
  {\path{doi:1000-6516(2016)03-0017-06}}.

\bibitem{patent}
T.~Zhang, J.~{Liu}, A pressure precision control device for dual-phase
  detector, Chinese Patent 202222807631.7.

\end{thebibliography}

\end{document}